# Direct-V2X Support with 5G Network-based Communications: Performance, Challenges and Solutions

M.C. Lucas-Estañ, *Member*, *IEEE*, B. Coll-Perales, *Member*, *IEEE*, T. Shimizu, *Member*, *IEEE*, J. Gozalvez, *Senior Member*, *IEEE*, T. Higuchi, *Member*, *IEEE*, S. Avedisov, *Member*, *IEEE*, O. Altintas, *Member*, *IEEE*, M. Sepulcre, *Senior Member*, *IEEE*

***Abstract*— This study analyzes the feasibility of supporting critical V2X services using 5G network-based Vehicle-to-Network-to-Vehicle (V2N2V) communications. The study evaluates the end-to-end latency of 5G V2N2V communications under different network deployments in single and multi-operator scenarios. The study shows that critical V2X services can be supported using 5G V2N2V communications over MEC-based network deployments. However, this requires the use of local peering points and shared data centers or MEC federation to address challenges arising from asymmetric network deployments. This opens the possibility for V2N2V communications to complement direct Vehicle-to-Vehicle (V2V) connections for increased reliability or to offload traffic under sidelink network congestion.**

## I. INTRODUCTION

5G and beyond networks can support Vehicle to Everything (V2X) services with direct Vehicle-to-Vehicle (V2V) communications over the PC5 interface, and/or with network-based Vehicle-to-Network (V2N) and Vehicle-to-Network-to-Vehicle (V2N2V) communications over the Uu interface (Fig. 1). Critical V2X services have been traditionally envisioned over direct or sidelink connections for reduced latency. However, the flexibility and features introduced with 5G can reduce the end-to-end (E2E) latency and open the possibility to support critical V2X services with network-based communications. For example, the 3GPP TR 37.910 shows that radio latency values below 2 ms can be achieved in uplink and downlink cellular connections in a range of RAN configurations (FDD or TDD frame structure and different numerologies and slot formats). Recent V2N trials in dedicated pilots under limited and controlled scenarios have reported mean 5G V2N latency values as low as 7.8 ms when using 5G networks with MEC for hosting the V2X application server [1].

The use of network-based communications has several benefits. For example, the network can provide controlled quality network-based connections and can opportunistically offload data traffic from congested V2V networks. In addition, the network can help connect vehicles using different technologies (like currently done with cellphones), and hence facilitate the deployment roadmap considering the longer life span of vehicles and the challenge to retrofit vehicles with new V2X technologies. In this context, V2N2V communications have the potential to complement (not necessarily replace) sidelink V2V communications. To this aim, it is necessary to demonstrate that V2N2V communications can support and scale critical V2X services with low latency requirements.

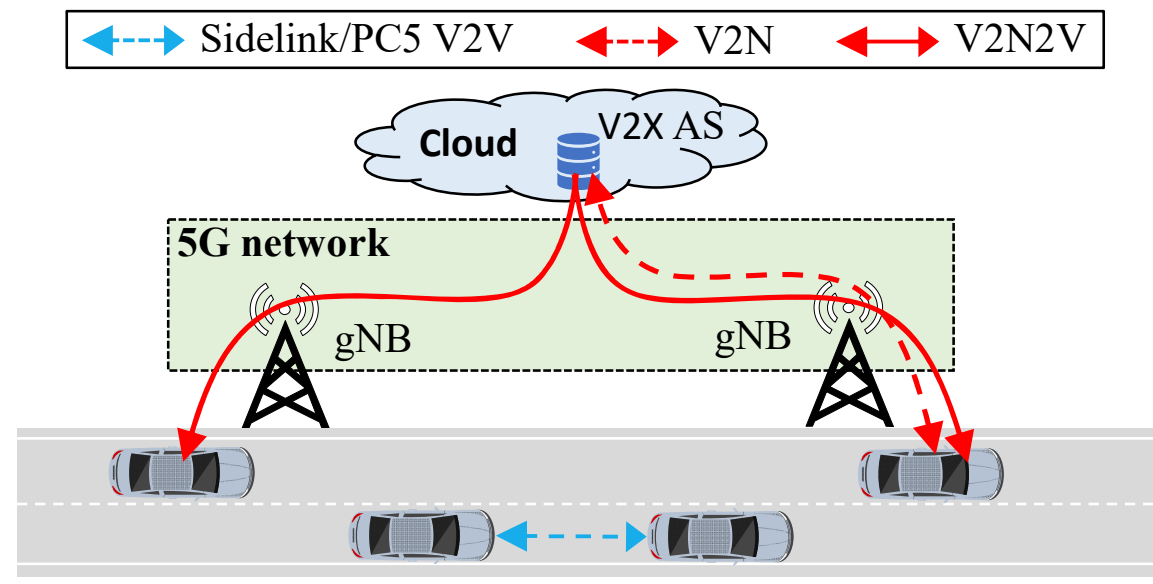

**Fig. 1.** V2X communication.

The latency (and jitter) of network-based communications depends on the network deployment and radio configuration, among others. This study analyzes the possibility to support critical V2X services using 5G network-based communications under centralized and MEC-based network deployments as illustrated in Fig. 1. Centralized networks deploy the V2X Application Server (AS) in a central cloud, while MEC-based networks deploy the V2X AS at the network edge to reduce and control the latency. In this context, this study makes several key contributions related to the support of direct V2X with 5G network-based communications. It shows that critical V2X services can be supported using 5G V2N2V communications

This work was supported in part by MCIN/AEI/10.13039/501100011033 (grants IJC2018-036862-I, PID2020-115576RB-I00) by the "European Union NextGenerationEU/PRTR" (TED2021-130436B-I00), and by Generalitat Valenciana (CIGE/2021/096). *Corresponding author: M.C. Lucas-Estañ.*

M.C. Lucas-Estañ, B. Coll-Perales, J. Gozalvez and M. Sepulcre are with Universidad Miguel Hernández de Elche, Spain. Contact emails: {m.lucas, bcoll, j.gozalvez, msepulcre}@umh.es.

T. Shimizu, T. Higuchi, S. Avedisov, and O. Altintas are with InfoTech Labs, Toyota Motor North America R&D, Mountain View, CA, U.S.A. Contact emails: {takayuki.shimizu, takamasa.higuchi, sergei.avedisov, onur.altintas}@toyota.com.

over MEC-based deployments in scenarios where vehicles are supported by a single Mobile Network Operator (MNO). The study also highlights and quantifies the challenges to support critical V2X services over 5G V2N2V communications when vehicles are served by different MNOs. The challenges are especially relevant when MNOs have asymmetric network deployments, e.g., when an MNO locates the V2X AS at a central cloud and the other one at a MEC system. The study then shows that MEC federation and shared data centers are viable solutions to address these challenges and maintain low latency E2E V2N2V connections in multi-MNO scenarios.

## II. 5G V2X Network-based Communications

### A. 5G system and network deployments

In 5G V2N2V communications, V2X data is transmitted from/to a vehicle through the gNB, the multiplexing nodes of the Transport Network (TN), and one or more User Plane Functions (UPFs) in the Core Network (CN) to/from the Data Network (DN) where the V2X AS is located. The V2X AS hosts the V2X services. The UPF that provides access to the DN is known as Packet Data Unit (PDU) Session Anchor (PSA) UPF. 5G can implement the PSA UPF at different locations (Fig. 2) and support different network deployments that impact the availability to support V2X services over 5G.

Fig. 2 depicts various network deployments. In centralized deployments, the V2X AS is in a central cloud and can leverage powerful computing and storage resources. However, data packets must traverse through the gNB, the TN, a chain of UPFs, and the Internet to reach the V2X AS. This increases the E2E latency and jitter, which can potentially impact the support of critical V2X services. Alternatively, the V2X AS can be placed on a MEC, and reduce the E2E latency. The MEC can be deployed, for example, at the CN ('MEC@CN' in Fig. 2). In this case, data packets travel through the gNB and the TN to the CN, where a PSA UPF steers the traffic towards the MEC. Alternatively, the MEC and (local) PSA UPF can be collocated with a multiplexing node of the TN (e.g., M1 in Fig. 2, 'MEC@M1') or the gNB ('MEC@gNB'). In MEC-based deployments, the data is processed closer to the vehicles than in centralized deployments. This reduces the E2E latency, the load on the TN and CN, the amount of data traffic that the V2X AS processes and therefore the processing power required for the V2X AS. However, deployed MEC nodes closer to the vehicles, augments the number of MEC nodes necessary to support V2X services ubiquitously.

### B. Mobility and QoS support

MEC-based deployments increase the probability of changing the PSA UPF as vehicles move, and 5G must support user mobility and service continuity with V2N2V communications. 5G Standalone (SA) introduces Session and Service Continuity (SSC) mode 3, which guarantees service continuity as a vehicle changes the serving PSA UPF and MEC. With SSC mode 3, 5G can establish a session with a new PSA UPF before the connection with the previous PSA UPF is released.

5G dynamically manages the Quality of Service (QoS) of V2X communications over the Uu using QoS flows. QoS flows are characterized by a set of QoS parameters such as Packet Delay Budget (PDB), Packet Error Rate (PER), and priority level. The values for the QoS parameters are indicated using a 5G QoS Identifier (5QI), selected from a set of standardized 5QIs [2]. Additionally, 5G can create network slices to support V2X services with specific attributes (e.g., delay tolerance, throughput, or area of service [3]). Network slicing allows for the implementation of tailor-made functionalities and network operation adaptation to the specific QoS needs of V2X services. It can also create dedicated logical network partitions isolated (and protected) from those used to support other services.

### C. 5G New Radio

The 5G QoS is strongly conditioned by the radio interface. 5G defines a highly flexible New Radio (NR) interface that can be configured to satisfy the requirements of the services it supports. At the physical layer, 5G NR defines several OFDM numerologies $\mu$=0,...,4 with a subcarrier spacings (SCS) equal to $15 \cdot 2^{\mu}$ kHz and slot durations from 1 ms to 0.0625 ms. 5G NR also defines different Modulation and Coding Schemes (MCSs) that trade-off spectrum efficiency and transmission reliability.

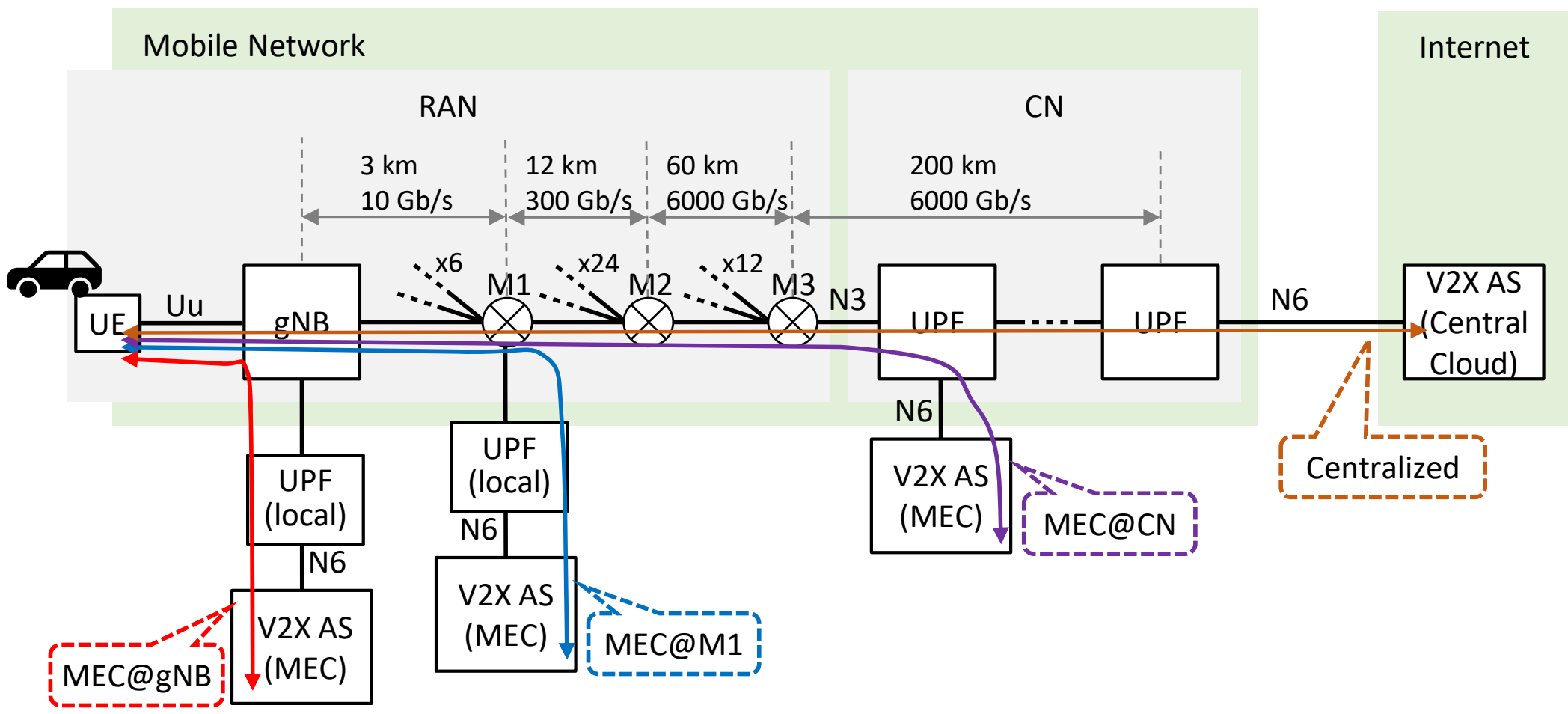

**Fig. 2.** 5G network deployments considered in this paper (arrows mark the path followed by the V2N2V data traffic in each deployment).

Asynchronous Hybrid Automatic Repeat Request (HARQ) or *k*-repetitions can improve the reliability at the MAC. HARQ requires feedback to request retransmission of a packet, which increases latency. Latency can be reduced using *k*-repetitions that transmits *k* replicas of the same packet in consecutive slots. However, this latency gain is achieved at the cost of a lower spectrum efficiency if unnecessary repetitions are transmitted. Transmissions at the MAC can be scheduled using dynamic or semi-static scheduling. Dynamic scheduling assigns new resources for each transmission. This impacts the latency due to the exchange of control messages between vehicles and the gNB to request/assign resources. The semi-static scheduling reduces the signaling as it pre-assigns radio resources periodically to vehicles. This approach reduces latency but can negatively impact the spectrum efficiency if pre-assigned resources are not used.

Broadcast and multicast communications are critical for the scalability of V2N and V2N2V communications. 3GPP Release 17 standards have introduced Radio Access Network (RAN) procedures and architectural enhancements for the support of multicast and broadcast services. V2X communications over the Uu are still only unicast in Release 17. Multicast/broadcast support is under study in Release 18 [4].

## III. REFERENCE SCENARIO AND LATENCY MODELLING

This work considers that the hierarchical TN architecture proposed by the ITU-T [5] is used to interconnect the RAN and the CN in all network deployments analyzed. This architecture considers 3 multiplexing nodes (Fig. 2). M1 multiplexes the traffic from 6 gNBs, while M2 and M3 multiplex the traffic from 24 M1 and 12 M2 nodes, respectively [5]. The distance and link capacity between nodes are also represented in Fig. 2. We estimate and reserve the fraction of the link capacities for V2X needed to prevent a backlog of V2X packets at TN or CN nodes. We dimension the processing power of the V2X AS to prevent any backlog following [6]. The number of processors required per MEC depends on the traffic generated by vehicles and the location of the MEC since the location impacts the aggregated traffic that an AS deployed at the MEC must handle. We follow the 3GPP guidelines for the evaluation of V2X [7], and consider a 3-lane highway with vehicle densities ranging from 10 to 80 veh/km/lane. Vehicles generate 10 or 50 pkt/s. This results in each gNB receiving V2X packets at a rate $\lambda$ ranging from 1040 to 41600 pkt/s. We consider the Cooperative Lane Change (CLC) service as a case study. According to 3GPP, CLC requires 90 and 99.99% of packets to be received in less than 25 and 10 ms, respectively, for vehicles with low (up to SAE level 3 [8]) or high (SAE level 4 or 5) Level of Automation (LoA) [9]. Following 3GPP standards [2], we utilize 5QI values 83 and 86 for the CLC service with low and high LoA respectively. We utilize MCSs with high error protection (target Block Error Rate -BLER- of $10^{-5}$) to support CLC with high LOA, and MCSs with lower error protection (target BLER of 0.1) to support CLC with low LoA. The RAN operates using Frequency Division Duplex (FDD), a 30 kHz SCS, full-slot transmissions, a default cell bandwidth of 20 MHz, and semi-static scheduling [9].

We use the models derived in [10] and [6] to estimate the E2E latency ($l_{E2E}$) of 5G V2N2V connections. $l_{E2E}$ accounts for the latency experienced at the radio ($l_{radio}$), transport ($l_{TN}$) and core ($l_{CN}$) networks, the Internet connection between the PSA UPF and V2X AS ($l_{UPF\text{-}AS}$), and the processing latency at the V2X AS ($l_{AS}$). For multi-MNO scenarios, $l_{E2E}$ must also consider the latency introduced in the peering points between the networks ($l_{pp}$). $l_{E2E}$ is expressed as:

$$l_{E2E} = l_{radio} + l_{TN} + l_{CN} + l_{UPF\text{-}AS} + l_{AS} + l_{pp}. \quad (1)$$

$l_{radio}$ is determined using the model in [10] and considering the reference scenario. $l_{TN}$ and $l_{CN}$ are computed following [6] as the sum of the propagation and transit delays over the TN and CN, respectively. Propagation delay represents the time packets need to travel through the links at the TN or CN and depends on the link distance. Transit delay accounts for the time needed to receive, process, and transmit packets at TN or CN nodes. We compute the transit delay using queueing theory, and it depends on the number of nodes packets pass through, network load, and link capacities. For the centralized deployment, $l_{UPF\text{-}AS}$ only intervenes in the centralized deployment. We model $l_{UPF\text{-}AS}$ using the empirical study in [11] that characterizes the round-trip time observed between source-target Internet nodes in the same country. We estimate $l_{pp}$ using empirical measurements in [12]. Peering points can be remote or local. Remote peering points are established at common Internet eXchange points (IXP). Local peering points can be direct links (e.g., fiber links), controlled Internet Protocol (IP) connections or managed wide area network connections subject to Service Level Agreements (SLA) between the MNOs. We estimate $l_{AS}$ using [13] and assume that the V2X AS only forwards the received packets. Table I reports the average, $90^{th}$ and $99.99^{th}$ percentile latency values for each latency component of a V2N2V connection following (1)[1]. Table II reports the average and $90^{th}$ and $99.99^{th}$ percentile values of the E2E latency of 5G V2N2V connections when all vehicles are served by the same MNO (single-MNO) or by multiple MNOs (multi-MNO). The values reported are valid for scenarios where vehicles transmit a packet to the V2X AS, and the V2X AS forwards the packet to a vehicle using a unicast transmission, or to several vehicles using a broadcast/multicast transmission[2].

[1] 3GPP establishes different reliability requirements for the CLC service based on the level of automation. 3GPP sets the reliability requirement to the $90^{th}$ percentile for the CLC service with low LoA, and to the $99.99^{th}$ percentile for the CLC service with high LoA. In this context, we use the highest error protection MCSs for the CLC service with high LoA and configure the CLC service with low LoA with lower error protection MCSs given its lower reliability requirement.

[2] Supporting V2X services with network-based communications over the Uu cannot scale with the traffic density if multiple unicast transmissions must be done per packet. For example, the $99.99^{th}$ percentile of the radio latency under $\lambda$=4150 pkt/s is equal to 3.1 ms if a packet is forwarded to a set of target vehicles

TABLE I

ROUND-TRIP LATENCY (IN MS) FOR THE LATENCY COMPONENTS OF A V2N2V CONNECTION

| Deployment / Latency component | Average | | | | 90th percentile | | | | 99.99th percentile | | | |
|---|---|---|---|---|---|---|---|---|---|---|---|---|
| | **Centralized** | **MEC @CN** | **MEC @M1** | **MEC @gNB** | **Centralized** | **MEC @CN** | **MEC @M1** | **MEC @gNB** | **Centralized** | **MEC @CN** | **MEC @M1** | **MEC @gNB** |
| $l_{radio}$ | 1.50-14.20 | | | | 1.90-24.61 | | | | 2.60-29.30 | | | |
| $l_{TN}$ | 2.36-2.36 | 2.36-2.36 | 0.85-0.88 | 0.41-0.42 | 2.37-2.37 | 2.37-2.37 | 0.88-0.93 | 0.42-0.45 | 2.42-2.44 | 2.42-2.44 | 1.04-1.25 | 0.51-0.61 |
| $l_{CN}$ | 2.01 | <0.01 | <0.001 | <0.001 | 2.01 | <0.01 | <0.001 | <0.001 | 2.01 | <0.01 | <0.001 | <0.001 |
| $l_{UPF\text{-}AS}$ | 10.30 | | | | 21.00 | | | | 43.00 | | | |
| $l_{pp\text{-}remote}$ | 13.00 | | | | 29.87 | | | | 99.21 | | | |
| $l_{pp\text{-}local}$ | 0.31 | | | | 0.43 | | | | 1.49 | | | |
| $l_{AS}$ | 0.50 | | | | 0.70 | | | | 0.75 | | | |

TABLE II

E2E LATENCY (IN MS) UNDER SINGLE AND MULTI-MNO SCENARIOS AND DIFFERENT 5G NETWORK DEPLOYMENTS

| Scenario | Peering Point | Deployment | Average | 90th percentile | 99.99th percentile |
|---|---|---|---|---|---|
| Single MNO | N/A | Centralized | 16.7-29.4 | 28.0-50.7 | 50.8-77.5 |
| | | MEC@CN | 4.4-17.1 | 5.0-27.7 | 5.8-32.5 |
| | | MEC@M1 | 2.9-15.6 | 3.5-26.2 | 4.4-31.3 |
| | | MEC@gNB | 2.4-15.2 | 3.0-25.8 | 3.9-30.7 |
| Multi-MNO with symmetric deployments | Remote | Centralized | 29.7-42.4 | 57.8-80.5 | 150.0-176.7 |
| | | MEC@CN | 17.4-30.1 | 34.8-57.5 | 105.0-131.7 |
| | | MEC@M1 | 15.8-28.6 | 33.4-56.1 | 103.6-130.5 |
| | | MEC@gNB | 15.4-28.2 | 32.9-55.6 | 103.1-129.9 |
| | Local | Centralized | 17.0-29.7 | 28.4-51.10 | 52.3-79.0 |
| | | MEC@CN | 4.7-17.4 | 5.4-28.1 | 7.3-34.0 |
| | | MEC@M1 | 3.2-15.9 | 3.9-26.7 | 5.9-32.8 |
| | | MEC@gNB | 2.7-15.5 | 3.5-26.2 | 5.4-32.2 |
| Multi-MNO with asymmetric deployments | N/A | MEC@CN+Centralized | 15.7-28.4 | 27.0-49.7 | 49.8-76.5 |
| | | MEC federation (MEC@CN) | 4.7-17.4 | 5.4-28.1 | 7.3-34.0 |
| | | Shared data center | 5.0-17.7 | 5.8-28.5 | 8.8-35.5 |

The tables report the latency values for all the network deployments in Fig. 2. When latencies (E2E or component) are reported as a range, the minimum and maximum values correspond to the latency experienced under the lowest (1040 pkt/s) and highest (41600 pkt/s) network traffic loads analyzed.

## IV. V2N2V LATENCY IN SINGLE-MNO SCENARIOS

MNOs traditionally rely on centralized deployments for provisioning mobile services. However, Table II shows that centralized deployments in single-MNO scenarios cannot support critical V2X services using V2N2V communications due to the latency introduced by the Internet connection between the CN and the AS in a central cloud ($l_{UPF\text{-}AS}$, Table I). Table II shows that the minimum E2E latency that can be guaranteed for 90% and 99.99% of the packets (28 and 50 ms, respectively) with centralized deployments is significantly higher than the latency required to support the CLC service with low and high LoA (25 and 10 ms, respectively).

Table II shows that MEC-based deployments reduce the E2E latency of V2N2V connections. The location where the MEC is deployed (RAN, TN, or CN) only lightly impacts the E2E latency. The results in Table II confirm that MEC-based deployments can support the CLC service with low LoA with a cell bandwidth of 20 MHz, except for the highest network load analyzed ($\lambda$=41600 pkt/s). Table I and Fig. 3 show that supporting the service under high loads is limited by the increase of the radio latency ($l_{radio}$). Fig. 3 depicts the 90$^{th}$ and 99.99$^{th}$ percentiles of the round-trip radio latency for the CLC service with low and high LoA, respectively[1]. Fig. 3 reports latency values for different cell bandwidths under medium and high network loads. We should note that the CLC service with high LoA is configured with a more robust 5G NR configuration (MCSs with highest error protection) than the service with low LoA following 3GPP requirements. The use of a more robust configuration consumes more radio resources, and explains the differences between the 90$^{th}$ and 99.99$^{th}$ percentile values (Fig. 3). Fig. 3 shows that the 90$^{th}$ percentile of the radio latency is well below the CLC requirement with low LoA (25 ms) when $\lambda$=10400pkt/s, and increases to nearly 25 ms when $\lambda$ increases to 41600 pkt/s and the bandwidth is 20 MHz. It is then not possible to support the CLC service under the highest load since the 90$^{th}$ percentile of E2E latency for any MEC-based deployment (Table II) exceeds 25 ms. The 90$^{th}$ percentile of the radio latency decreases drastically even under using multicast/broadcast communications and the bandwidth is 20 MHz. If the packet must be forwarded to 4 or more vehicles using unicast transmissions, this latency increases beyond the 10 ms requirement established to support the CLC service with high LoA.

the highest load (2.6 ms) if we increase the bandwidth to 30 MHz, and it is again possible to support the CLC service. For example, the 90th percentile of the E2E latency for the 'MEC@CN' deployment drops to 5.7 ms.

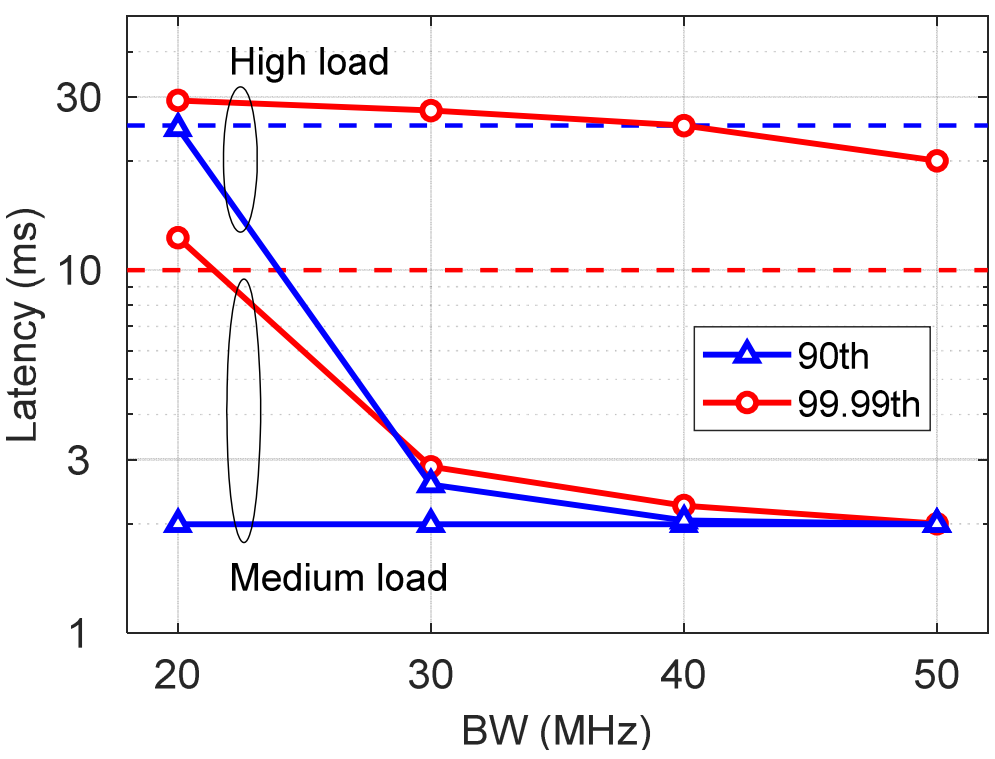

**Fig. 3.** 90th and 99.99th percentiles of the round-trip radio latency ($l_{radio}$) as a function of the cell bandwidth for medium ($\lambda$=10400 pkt/s) and high ($\lambda$=41600 pkt/s) loads. $\lambda$=10400 pkt/s corresponds to a vehicular density of 20 or 50 veh/km/lane with vehicles generating 50 or 20 pkt/s. $\lambda$=41600 pkt/s corresponds to a vehicular density of 80 veh/km/lane with vehicles generating 50 pkt/s. $l_{radio}$ is independent of the network deployment.

Table II shows that MEC-based deployments can support the CLC service with high LoA under low loads. However, this is not possible at medium loads ($\lambda$=10400pkt/s) with 20 MHz since the radio latency overpasses the 10 ms requirement (Fig. 3). We can support the service under medium loads if we increase the bandwidth to 30 MHz since the 99.99th percentile of the radio latency decreases to 2.9 ms (Fig. 3). This results in the 99.99th percentile of the E2E latency for the 'MEC@CN' deployment dropping down to 6.1 ms. However, we cannot support the CLC service with high LoA under the highest load ($\lambda$=41600 pkt/s) even if bandwidth increases to 50 MHz. This is due to the use of a more robust 5G NR configuration (MCSs with highest error protection) to support CLC with high LoA that consumes more radio resources. Our evaluation reveals that we can support the CLC service with high LoA up to a load of 31200 pkt/s if we increase the bandwidth to 50 MHz. In this case, the 99.99th percentile of the round-trip radio and E2E latency (for 'MEC@CN') are equal to 6.6 and 9.8 ms, respectively.

## V. V2N2V Latency in Multi-MNO Scenarios

The previous section considers that two vehicles communicating over a V2N2V connection are served by the same MNO. However, different MNOs may serve each vehicle which creates additional challenges. These include increased latency due to peering point connections between the networks, as well as the impact of asymmetric network deployments and configurations.

### A. Interconnection between networks

Two vehicles served by different MNOs could establish a V2N2V connection using a central cloud where, for example, a third party hosts the V2X AS. This requires an Internet connection to the cloud, but this connection significantly increases the latency (Table I). This makes not possible to support critical V2X services such as CLC. An alternative is to interconnect centralized network deployments through remote IXP. Nevertheless, these remote connections add high latencies ($l_{pp\text{-}remote}$ in Table I) that prevent centralized deployments to support the CLC service with low or high LoA. Table II shows that the minimum E2E latency guaranteed to 90% and 99.99% of the packets is 57.8 and 150 ms, respectively, which exceeds the CLC requirements. Consequently, we focus the rest of this section on MEC-based deployments.

MNOs can interconnect their networks using local peering points between their MECs or using remote peering points (through IXP) between the PSA UPFs that serve the MECs. The use of remote peering points increases the E2E latency of V2N2V communications in all MEC-based deployments (Table II and $l_{pp\text{-}remote}$ in Table I). In this case, the CLC service cannot be supported, regardless of the level of automation. It is possible to support the CLC service for all MEC-based deployments using local peering points under low and medium network loads but not under the highest load (Table II). This is due to the significant increase in radio latency with the network load ($l_{radio}$ in Table I). We can reduce $l_{radio}$ and the E2E latency by increasing the cell bandwidth. For example, the E2E latency guaranteed to 90% of the packets decreases from 28.1 to 6.1 ms if we increase the cell bandwidth to 30 MHz under the 'MEC@CN' deployment using local peering points and the highest network load. Local peering points between networks then represent a good option to control the E2E latency and support critical V2X services in multi-MNO scenarios. However, deploying local peering points requires significant investments from MNOs and new SLAs between MNOs, which increases network management complexity.

### B. Asymmetric network deployments

The multi-MNO scenarios analyzed so far consider symmetric deployments where both MNOs use the same network deployment and configuration. However, this may not always be the case, and asymmetric deployments present additional challenges for V2X support across MNOs. For example, let's suppose that two vehicles are supported by different MNOs. The first MNO opts for a centralized deployment with the V2X AS located at a central cloud, while the second one deploys a MEC at the CN. In this case, the V2X traffic generated by a vehicle supported by the first MNO must reach the central cloud through the Internet, and the traffic is then forwarded to the second network (through the Internet) to reach its MEC. This scenario increases the E2E latency compared to a symmetric deployment. In this asymmetric deployment and under the lowest load, the 90th and 99.99th percentile E2E latency values increase to 27 and 49.6 ms, respectively ('MEC@CN+Centralized' in Table II), compared to 5.4 and 7.3 ms with a symmetric 'MEC@CN' deployment with local peering points. Therefore, the CLC service cannot be supported in this asymmetric scenario despite one of the operators deploying a MEC.

## VI. MEC Federation and Shared Data Centers

Network sharing is a trend in the cellular industry that accelerates deployments, reduces cost, and improves network quality. Standards and industry organizations (e.g., 3GPP [2], ETSI [14] and 5GAA [15]) have recently proposed to extend the concept of sharing to computing resources through MEC federation and shared data centers. These proposals can help address existing challenges in multi-MNO scenarios.

A MEC federation enables the shared usage of MEC services and resources hosted in different networks [14]. With MEC federation, vehicles can access services and computing resources hosted on their own MNO's MEC and on the MEC of a different MNO. This allows for access to the federated MEC system with the best service level, and even enables the implementation of compute-intensive services at the edge through collaboration between MEC systems. MEC federation also helps support service continuity when a vehicle moves out of its MNO's service area. Alternatively, MNOs can deploy their MEC and AS in a shared data center outside their domains. MNOs connect their networks to the shared data center using UPF nodes collocated at the center [15]. Shared data centers offer similar benefits to MEC federation. For example, vehicles can access resources supported by another MNO's MEC in the shared data center. A single MEC can be used, for example, to coordinate a maneuver between two vehicles served by different MNOs through a V2N2V connection. This reduces the latency compared to a scenario without MEC federation or shared data centers where the V2X AS of the MNOs need to exchange data to coordinate the maneuver.

MEC federation and shared data centers facilitate the communication between vehicles supported by different MNOs and reduce the E2E latency of V2N2V communications. This is particularly useful to address the challenges resulting from asymmetric deployments as illustrated in Fig. 4. Fig. 4 depicts a scenario where MNOs with asymmetric network deployments support two communicating vehicles. MNO A deploys a MEC at the CN hosting the V2X AS, while MNO B deploys the V2X AS on the central cloud. Without MEC federation or shared data centers, a packet transmitted by a vehicle served by MNO B must be routed and processed at the AS on the central cloud before being sent through MNO A's network and its MEC to the receiving vehicle (path labeled as 'w/o MEC federation' in Fig. 4-a). On the other hand, if both networks use MEC federation, vehicles served by MNO B can directly access the V2X AS available at MNO A's MEC. In this case, the path packets follow between vehicles A and B (labeled 'w/ MEC federation' in Fig. 4-a) is significantly reduced compared to the original asymmetric scenario as the packet does not need to be routed over the Internet to a central cloud. This is also avoided when the MNOs deploy their MECs at a shared data centers to which they connect using a controlled local peering point connection (Fig. 4-b).

The impact of MEC federation and shared data centers on the E2E latency of V2N2V communications is visible in Table II. The table shows that the 90$^{th}$ and 99.99$^{th}$ percentiles of the E2E latency under the lowest load exceeds the requirements of the CLC service if we do not utilize MEC federation or shared data centers ('MEC@CN+Centralized' in Table II). With MEC federation, the 90$^{th}$ and 99.99$^{th}$ percentiles of the E2E latency decrease to 5.4 and 7.3 ms respectively under the lowest load ('MEC federation (MEC@CN)' in Table II). In this case, it is possible to support the CLC service with low and high LoA. The E2E latency can also be reduced if both MNOs deploy their MECs in a shared data center, and the networks are connected to the center using a controlled local peering point connection (Fig. 4-b). In this case, the 90$^{th}$ and 99.99$^{th}$ percentiles of the E2E latency are reduced to 5.8 and 8.6 ms, respectively, under the lowest load ('shared data center' in Table II).

MEC federation and shared data centers provide similar benefits to address challenges in multi-MNO scenarios but differ in implementation complexity. With MEC federation, all MNOs that participate in the federated MEC system need to establish local peering point connections between their networks and the relevant MECs of all the MNOs to guarantee low latencies. The use of a shared data center only requires such type of connection between the CNs and the UPFs in the shared data center. Shared data centers can represent a more cost-effective and less complex solution to support network-based V2N2V communications in multi-MNO scenarios. This does not mean that MNOs should not deploy their own MECs. These MECs can be a better option when considering V2N services supported by a single MNO or other non-V2X services that rely exclusively on the MNO in question.

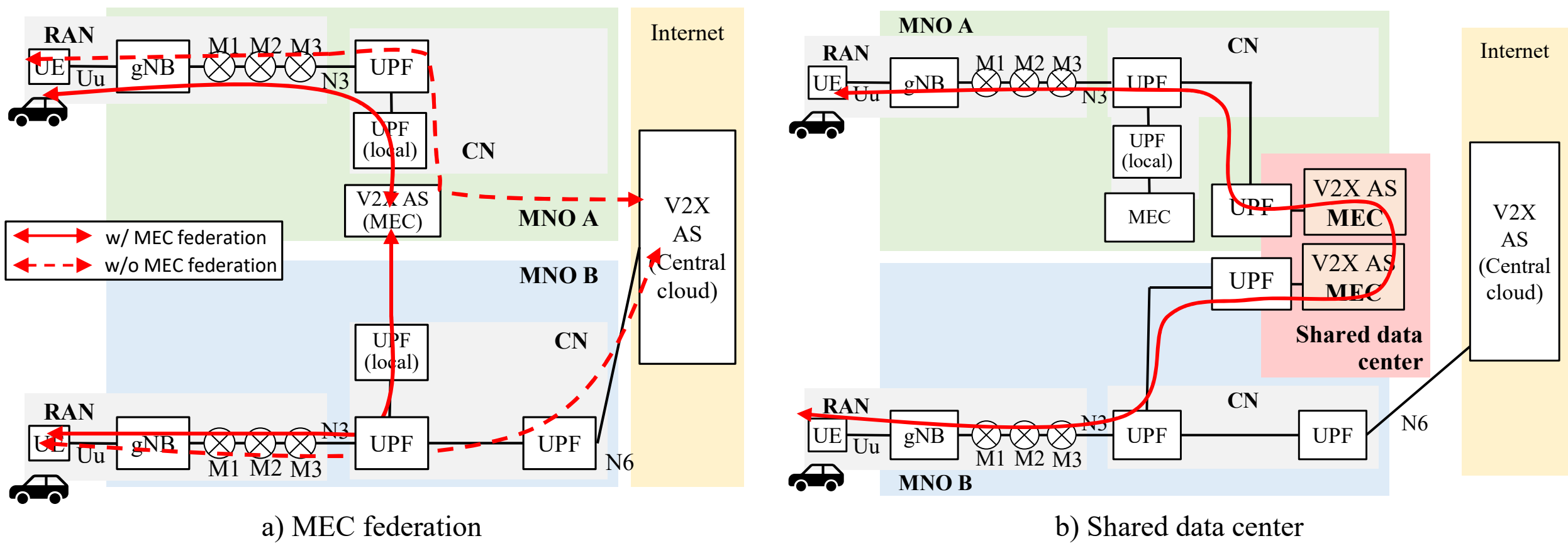

**Fig. 4.** 5G V2N2V communications in multi-MNO scenarios with asymmetric network deployments.

## VII. Conclusions and Discussion

This study has demonstrated that critical V2X services can be supported using 5G network-based V2N2V communications over MEC-based network deployments and under certain conditions. This opens the possibility to utilize V2N2V communications for services initially designed to operate over direct V2V connections. V2N2V connections should not necessarily replace direct V2V connections but can complement them. For instance, V2N2V connections can be used as redundant links to increase reliability or to offload some traffic when the V2V network is congested.

The capacity to support critical V2X services with V2N2V communications depends on the network deployment, network load, bandwidth, radio configuration, and service requirements. Some of the main challenges arise when vehicles are served by different MNOs. In this scenario, local peering points or controlled connections between networks are necessary to guarantee E2E latency and QoS requirements. However, this incurs into higher deployment costs and management complexity. Additionally, the complexity of control processes and signaling increases when establishing inter-MEC system communications (for example, to discover MEC platforms that belong to different systems).

Several standardization and industry groups are addressing the challenges for V2X support and continuity in multi-MNO scenarios. For example, ETSI is developing interfaces and mechanisms for MEC federation implementation [14], while the MEC4Auto working group of 5GAA is considering shared data centers for multi-MNO scenarios [15]. MEC federation and shared data centers offer significant benefits in terms of management complexity, scalability, and deployment costs. Both solutions enable MNOs to collaborate and coordinate their deployments as V2X services are gradually rolled-out. The 5G Future Forum (5GFF) promotes collaboration among MNOs with interoperable 5G MEC solutions through an MNO-agnostic Application Programming Interface (API). This API allows MNOs to expose their 5G and MEC capabilities, and end-users to discover the best MEC to connect to. GSMA (Global System for Mobile Communications Association) promotes collaboration between MNOs through an Operator Platform (OP)[3]. The OP is based on a federated model and defines a common platform exposing 5G operator services/capabilities to customers/developers. The OP aims to provide end-users served by the OP with the same service level as if they were supported by a single MNO. The first phase focuses on the edge, and GSMA plans to expand in future phases with connectivity and network slicing. Network slicing in 5G SA plays an important role in guaranteeing V2X requirements, relying on standardized Slice/Service types (SST) that define the expected behavior of a network slice [2]. Supporting the service in an end-to-end connection involving several MNOs will require SLAs among the MNOs. Defining and implementing these SLAs are not straightforward technical and organizational tasks, especially with asymmetric deployments where MNOs cannot support the same capabilities and features.

[3] GSMA's Operator Platform: https://www.gsma.com/futurenetworks/5g-operator-platform/.

## BIOGRAPHIES

**M. Carmen Lucas-Estañ** received her M.Sc. and Ph.D. degrees in Telecommunications Engineering from the Universidad Miguel Hernandez (UMH) de Elche, Spain. She is currently a Part-time Lecturer and Research Fellow at the UMH. She received the Santander-UMH Award for Young Researchers in 2021 and 2022. She is currently working on 6G/B5G networks, industrial wireless networks, and V2X communications. She was Guest Editor of a Special Issue in the Wireless Communications and Mobile Computing magazine.

**Baldomero Coll-Perales** received his M.Sc. and Ph.D. degrees in Telecommunications Engineering from the Universidad Miguel Hernandez de Elche (UMH), Spain. He is a Juan de la Cierva Research Fellow at UMH working on advanced mobile and wireless communication systems applied to critical verticals. He has hold visiting research positions at UPCT (Spain), Hyundai Motor Europe (Germany), IIT-CNR (Italy), WINLAB-Rutgers University (USA) and King's College London (UK). He serves/has served as Associate Editor for the International Journal of Sensor Networks and Springer's Telecommunication Systems, and as Track Co-Chair for IEEE VTC-Fall 2018.

**Takayuki Shimizu** received the B.E., M.E., and Ph.D. degrees from Doshisha University, Kyoto, Japan, in 2007, 2009, and 2012, respectively. From 2009 to 2010, he was a Visiting Researcher at Stanford University, CA, USA. From 2012 to 2019, he worked at TOYOTA InfoTechnology Center, U.S.A., Inc. Currently, he is a Senior Principal Researcher at Toyota Motor North America, Inc., R&D InfoTech Labs, where he works on the research and standardization of wireless vehicular communications. He received the 2010 TELECOM System Technology Award for Student from the Telecommunications Advancement Foundation and the 2020 IEEE Vehicular Networking Conference Best Paper Award.

**Javier Gozalvez** is a Full Professor at the Universidad Miguel Hernández de Elche (UMH) in Spain, where he leads research activities in vehicular networks, 6G/B5G, and industrial wireless networks. He currently serves as the Editor-in-Chief of the IEEE Vehicular Technology Magazine and is an elected member of the Board of Governors of the IEEE Vehicular Technology Society (IEEE VTS). He was the President of the IEEE VTS in 2016 and 2017. He holds a degree in electronics engineering from the Engineering School ENSEIRB (Bordeaux, France), and a PhD in mobile communications from the University of Strathclyde in Glasgow, U.K.

**Takamasa Higuchi** is a Principal Researcher at InfoTech Labs, Toyota Motor North America R&D. He received his B.E., M.E. and Ph.D. degrees from Osaka University, Japan, in 2010, 2012, and 2014, respectively. He started his professional carrier as an Assistant Professor at Osaka University, where he led research projects on mobile computing and networking. He joined Toyota in 2016 and has since been working on vehicular communications and vehicular cloud computing.

**Sergei S. Avedisov** received his Ph.D. degree in mechanical engineering from the University of Michigan, Ann Arbor, MI, USA, in 2019. He currently works as a principal researcher at Toyota R&D InfoTech Labs, Mountain View, CA, USA. His research interests include cooperative automated driving, cooperative perception, cooperative maneuvering, vehicle-to-everything communications, and cooperative platooning in mixed traffic.

**Dr. Onur Altintas** is the InfoTech Labs Fellow and a Senior Executive at InfoTech Labs, Toyota North America R&D, in Mountain View, California. He has been with the Toyota Group since 1999 in various roles in New Jersey, Tokyo, and California. He is the co-founder and has been the general co-chair of the IEEE Vehicular Networking Conference (IEEE VNC) since 2009. He serves as an associate editor for IEEE Vehicular Technology Magazine, IEEE ITS Magazine and IEEE Transactions on Intelligent Vehicles. He holds a Ph.D. from The University of Tokyo, and M.S. and B.S. from ODTÜ in Ankara, Türkiye, all in EE. He is an IEEE VTS Distinguished Speaker.

**Miguel Sepulcre** received the Telecommunications Engineering degree (2004) and the PhD degree in communications technologies (2010) from Universidad Miguel Hernández de Elche (UMH), Spain. He is an Associate Professor at UMH, and a member of UWICORE Research Laboratory working in V2X networks and industrial wireless networks. He was an Expert of the ETSI STF-585 on multi-channel operation. He was awarded by the COIT for the best national PhD thesis. He was TPC Co-Chair of IEEE VTC2018-Fall, IEEE/IFIP WONS 2018, and IEEE VNC 2016. He serves as an Associate Editor for IEEE Vehicular Technology Magazine and IEEE Communications Letters.